\documentclass[preprint,preprintnumbers,amsmath,amssymb]{revtex4}
\usepackage{amsfonts}
\usepackage{amssymb}
\usepackage{latexsym}
\usepackage{graphicx}% Include figure files
\newcommand{\al}{\alpha}

\newcommand{\be}{\beta}

\newcommand{\si}{\sigma}
\newcommand{\Si}{\Sigma}

\newcommand{\ga}{\gamma}

\newcommand{\De}{\Delta}

\newcommand{\rar}{\rightarrow}

\newcommand{\non}{\nonumber}
\begin{document}

%\preprint{M\'exico ICN-UNAM 01/06, \ June 2006 }

\title{The H$_3^+$ molecular ion in a magnetic field:
                   linear parallel configuration}

\author{A.~V.~Turbiner}
\email{turbiner@nucleares.unam.mx}
\author{N.~L.~Guevara}
\email{nicolais@nucleares.unam.mx}
\author{J.~C.~L\'opez Vieyra}
\email{vieyra@nucleares.unam.mx}

\affiliation{Instituto de Ciencias Nucleares, Universidad Nacional
Aut\'onoma de M\'exico, Apartado Postal 70-543, 04510 M\'exico,
D.F., Mexico}

\date{\today}

\begin{abstract}
A first detailed study of the ground state of the H$_3^+$ molecular
ion in linear configuration, parallel to a magnetic field direction,
and its low-lying $\Si,\Pi,\De$ states is carried out for magnetic
fields $B=0-4.414 \times 10^{13}\,$G in the Born-Oppenheimer
approximation. The variational method is employed with a single
trial function which includes electronic correlation in the form
$\exp{(\ga r_{12})}$, where $\ga$ is a variational parameter. It is
shown that the quantum numbers of the state of the lowest total
energy (ground state) depend on the magnetic field strength. The
ground state evolves from the spin-singlet ${}^1\Si_g$ state for
weak magnetic fields $B \lesssim 5 \times 10^{8}\,$G to a
weakly-bound spin-triplet ${}^3\Si_u$ state for intermediate fields
and, eventually, to a spin-triplet $^3\Pi_u$ state for $5 \times
10^{10}\,\lesssim B \lesssim 4.414 \times 10^{13}\,$G. Local
stability of the linear parallel configuration with respect to
possible small deviations is checked.

\end{abstract}

\pacs{31.15.Pf,31.10.+z,32.60.+i,97.10.Ld}

% physics/0606083

\maketitle

\section{\protect\bigskip introduction}

The behavior of atoms, molecules and ions placed in a strong
magnetic field has attracted a significant attention during the last
two decades (see, in particular, review papers
\cite{Liberman:1995,Lai:2001,Turbiner:2006}). It is motivated by
both pure theoretical interest and by possible practical
applications in astrophysics and solid state physics. In particular,
the knowledge of the energy levels can be important for
interpretation of the spectra of white dwarfs (where a surface
magnetic field ranges in $B\approx 10^{6}-10^{9}$\,G) and neutron
stars where a surface magnetic field varies in $B\approx
10^{12}-10^{13}$\,G, and even can be $B\approx 10^{14}-10^{16}$\,G
for the case of magnetars.

Recently, it was announced that in a sufficiently strong magnetic
field $B \gtrsim 10^{11}$\,G  the exotic molecular ion H$_3^{2+}$
can exist in linear configuration with protons situated along the
magnetic line \cite{Turbiner:1999} (for discussion see a review
\cite{Turbiner:2006}). In general, it is a metastable long-living
system which decays to H$_2^+ + p$. However, at $B \gtrsim
10^{13}$\,G the ion H$_3^{2+}$ becomes stable. This system does not
exist without or for weak magnetic fields. The ion H$_3^{2+}$
constitutes the simplest one-electron polyatomic molecular ion in a
strong magnetic field. The H$_3^{2+}$ ion has been proposed as being
the most abundant chemical compound in the atmosphere of the
isolated neutron star 1E1207.4-5209 \cite{Turbiner:2004m}. A
detailed review of the current status of one-electron molecular
systems, both traditional and exotic, that might exist in a magnetic
field $B \geq 10^{9}$\,G can be found in \cite{Turbiner:2006}.

The molecular ion $H_3^+$ is the simplest stable two-electron
polyatomic molecular ion. It has a long history since its discovery
by J.J.~Thomson \cite{thomson}. Its exceptional importance in
astrophysics related to interstellar media explains the great
interest in this ion from astronomy, astrophysics and chemistry
communities (for a detailed review, see, \cite{tennyson}). For all
these reasons, there have been extensive theoretical and
experimental works on this molecular ion since the pioneering
(semi-quantitative) work by Coulson \cite{coulson}.

The first variational calculations \cite{Hirschfelder} of the total
energy of the molecular ion H$_3^+$ showed that the equilibrium
configuration might be either linear or equilateral triangular.
However, this was not well-established until 1964
\cite{Christoffersen} when it was shown that the equilibrium
configuration for the state of the lowest total energy is an
equilateral triangular configuration, while the linear configuration
of the H$_3^+$ ion may occur in excited state(s). Since that time a
large number of excited states has been studied \cite{schaad} (for a
general review, see \cite{tennyson}). In particular, it has been
found that there exists a single spin-triplet state which appears in
a linear configuration ${}^3\Si_u$. This is also the unique known
state of H$_3^+$ in the linear configuration. No spin-triplet states
have been found for a triangular (spacial) configurations so far.

Although the molecular ion H$_3^+$ is characterized by the
equilateral triangular configuration as being the optimal in
field-free case, it is expected that in a magnetic field $B \approx
0.2$\,a.u. (see below) a linear configuration, parallel to a
magnetic field direction, gives the lower total energy and becomes
the optimal configuration. Somehow, a similar phenomenon already
happened for the one-electron exotic molecular ion H$_3^{2+}$
\cite{Turbiner:2002} where the optimal configuration is triangular
at $10^8 \lesssim B \lesssim 10^{11}\,$G and becomes linear parallel
at $B \approx 10^{11}\,$G. It is worth noting that for H$_3^+$ in
field-free case the difference between the total energy of the
ground state (triangular configuration) and of the lowest linear
configuration is very small, $\approx 0.13$\,Ry, in comparison to
characteristic energies in a magnetic field.

To the best of our knowledge there exists a single attempt to
explore the molecular ion H$_3^+$ in a magnetic field \cite{warke}.
We repeated all numerical calculations of this work following its
guidelines with use of its formulas (see below, Tables~I, V, VI) -
in fact, no single number from \cite{warke} was confirmed. However,
in \cite{warke} it was made a qualitative statement that with a
magnetic field increase the transition from equilateral stable
equilibrium configuration to linear equilibrium configuration may
occur. This statement we confirm. We predict that this transition
takes place at a magnetic field $\approx 0.2\,$a.u. A detailed study
of a triangular configuration and of this transition will be
published elsewhere \cite{tg}.

Atomic units are used throughout ($\hbar$=$m_e$=$e$=1), although
energies are expressed in Rydbergs (Ry). The magnetic field $B$ is
given in a.u. with a conversion factor $B_0 = 2.35 \times 10^9$\,G.

\section{Generalities}

\setcounter{figure}{0}

Let us consider a system of three protons and two electrons
$(pppee)$ placed in a uniform constant magnetic field. If for such a
system a bound state is developed it corresponds to the molecular
ion H$_3^+$. We assume that the protons are infinitely massive (the
Born-Oppenheimer approximation of zero order). They are situated
along the magnetic field direction forming a linear chain (we call
it ``the parallel configuration"). The Hamiltonian which describes
this system when the magnetic field is oriented along the $z$
direction, ${\bf B}=(0,0,B)$ is \footnote{The Hamiltonian is
normalized by multiplying on the factor 2 in order to get the
energies in Rydbergs}
\begin{equation}
 {\cal H} =\sum_{\ell=1}^2 \left( {\hat {\mathbf p}_{\ell}+{\cal A}_{\ell}}
 \right)^2 -\sum_{\buildrel{{\ell}=1,2}\over{\kappa =A,B,C}}
 \frac{2}{r_{{\ell}\, \kappa}}\,
 + \frac{2}{r_{12}}+ \frac{2}{R_+} + \frac{2}{R_-}
 + \frac{2}{R_++R_-} + 2{\bf{B}} \cdot {\bf S} \ ,
\end{equation}
(see Fig.~1 for the geometrical setting and notations), where ${\hat
{\mathbf p}_{\ell}}=-i \nabla_{\ell}$ is the 3-vector of the
momentum of the ${\ell}$th electron, the index $\kappa$ runs over
protons $A, B$ and $C$, $r_{12}$ is the interelectron distance and
$\bf{S}=\hat s_{1}+\hat s_{2}$ is the operator of the total spin.
${\cal A}_{\ell}$ is a vector potential which corresponds to the
constant uniform magnetic field $\bf B$. It is chosen to be in the
symmetric gauge,
\begin{equation}
   {\cal A}_{\ell}= \frac{1}{2}({\bf{B}} \times \ {\bf{r}}_{\ell})\
   =\ \frac{B}{2} (-y_{\ell},\ x_{\ell},\ 0)\ .
\end{equation}
Finally, the Hamiltonian can be written as
\begin{equation}
 %\hspace{-10pt}
  {\cal H} =\sum_{{\ell}=1}^2 \left(- {\mathbf\nabla}^2_ {\ell}
  +\frac{B^2}{4} \rho_{\ell}^2 \right) -
  \sum_{\buildrel{{\ell}=1,2}\over{\kappa =A,B,C}}
  \frac{2}{r_{{\ell}\, \kappa}}  + \frac{2}{r_{12}}+ \frac{2}{R_+} +
  \frac{2}{R_-} + \frac{2}{R_++R_-}+ B (\hat L_z +2\hat S_z )\  ,
\end{equation}
where  $\hat L_z=\hat l_{z_1}+\hat l_{z_2}$ and $\hat S_z=\hat
s_{z_1}+\hat s_{z_2}$ are the z-components of the total angular
momentum and total spin, respectively, and
$\rho_{\ell}=\sqrt{x_{\ell}^2+y_{\ell}^2}$.

%%%%%%%%%%%%%%  FIGURE:1  %%%%
%%%%%%%%%%%%%%  H3++ geometrical settings
\begin{figure}[tb]
\label{figg}
\begin{center}
   \includegraphics*[width=5.in,angle=0]{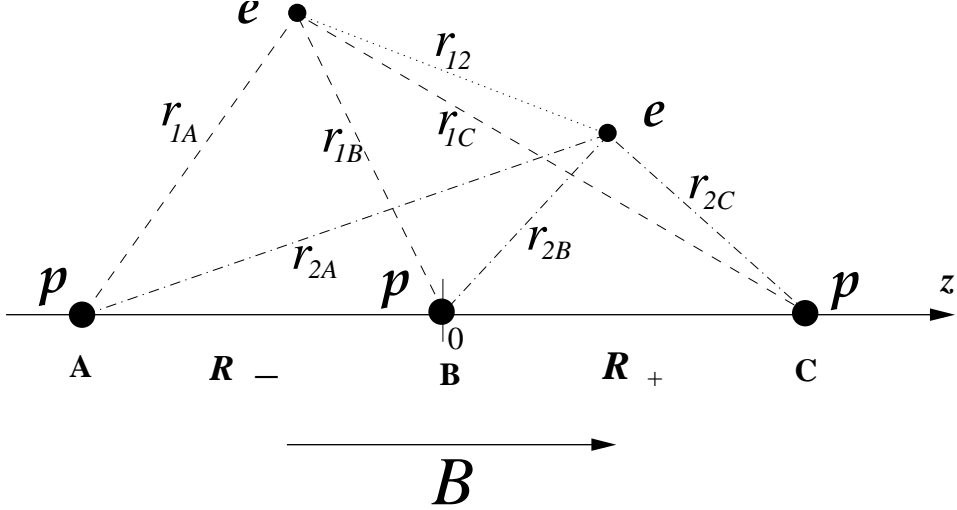}
    \caption{Geometrical setting  for the $H_3^{+}$
      ion in linear configuration parallel to a magnetic field directed
      along $z$-axis. The
      protons (marked by bullets) are situated on the $z$-line at
      distances $R_{\pm}$ from the central proton which is placed at
      the origin.}
\end{center}
\end{figure}

The problem under study is characterized by three conserved
quantities: (i) the operator of the $z$-component of the total
angular momentum (projection of the angular momentum on the magnetic
field direction) giving rise to the magnetic quantum number $m$,
(ii) the spatial parity operator $P({\vec r_1} \rar -{\vec
r_1},{\vec r_2} \rar -{\vec r_2})$ which has eigenvalues $p=\pm
1$(gerade/ungerade)
%(ii) the square of the total spin operator giving rise to the spin
%quantum number $S$ and
(iii) the operator of the $z$-component of the total spin
(projection of the total spin on the magnetic field direction)
giving rise to the total spin projection $m_s$. Hence, any
eigenstate has three explicit quantum numbers assigned: the magnetic
quantum number $m$, the total spin projection $m_s$ and the parity
$p$. For the case of two electrons the total spin projection $m_s$
takes values $0,\pm 1$.

As a magnetic field increases a contribution from the Zeeman term
(interaction of spin with magnetic field, ${\bf{B}} \cdot {\bf{S}}$)
becomes more and more important. It seems natural to assume that for
small magnetic fields a spin-singlet state is a state of a lowest
total energy, while for large magnetic fields it should be a
spin-triplet state with $m_s=-1$, where the electron spins are
antiparallel to the magnetic field direction ${\bf{B}}$. The total
space of eigenstates is split into subspaces (sectors), each of them
is characterized by definite values of $m$, $p$ and $m_s$.  It is
worth noting that the Hamiltonian $\cal H$ is invariant with respect
to reflections $z_1 \to -z_1$ and $z_2 \to -z_2$ ($z$-parity
operator $P_z$). Hence, any eigenstate is characterized by the
quantum numbers $\si_N=\pm 1$ for positive/negative $z$-parity (this
symmetry accounts for the interchange of the nuclei $A$ and $C$ if
they are situated symmetrically with respect to $B$).

In order to classify eigenstates we follow the convention widely
accepted in molecular physics using the quantum numbers $m, p$ and
the total spin $S$ without indication to the value of $m_s$.
Eventually, the notation is ${}^{2S+1} M_{p}$, where $2S+1$ is the
spin multiplicity which is equal to $1$ for spin-singlet state
($S=0$) and $3$ for spin-triplet ($S=1$), as for the label $M$ we
use Greek letters $\Si, \Pi, \De$ that mark the states with $|m|=0,
1, 2,...$, respectively, and the subscript $p$ (the spatial parity
quantum number) takes gerade/ungerade($g/u$) labels describing
positive $p=+1$ and negative $p=-1$ parity, respectively. There
exists a relation between the quantum numbers corresponding to the
$z$-parity (interchange of nuclei A and C) and the spatial parity:
\[
p=(-1)^{|m|}\ \si_N \,.
\]
Present consideration is limited to the states with magnetic quantum
numbers $m=0,-1,-2$ because the total energy of the lowest energy
state (the ground state) for any sector with $m>0$ is always larger
than anyone with $m \leq 0$.

As a method to explore the problem we use the variational procedure.
The recipe of choice of trial functions is based on physical
arguments \cite{turbinervar}. As a result the trial function for a
lowest energy state with magnetic quantum number $m$ is chosen in
the form
\begin{eqnarray}
\label{psi}
 \psi^{(trial)} &=&
       (1+\si_e P_{12})
       (1+\si_{\scriptscriptstyle N} P_{AC})
       (1+\si_{\scriptscriptstyle N_a} P_{AB}
         +\si_{\scriptscriptstyle N_a} P_{BC})\times
 \non
 \\
 \hspace{20pt}\rho_1^{\mid m \mid}e^{im\phi_1} \,\, e^{\gamma r_{12}}
  &&\hspace{-22pt}{e}^{-\al_1 r_{1A}-\al_2 r_{1B} -\al_3 r_{1C}-
  \al_4 r_{2A} -\al_5 r_{2B}-\al_6 r_{2C}
  -  B  \be_{1}\frac{\rho_1^2}{4}-
  B \be_{2}\frac{\rho_2^2}{4} }
\end{eqnarray}
where $\si_e=\pm 1$ stands for spin singlet (+) and triplet states
$(-)$, while $\si_N=1,-1$ stands for nuclear gerade and ungerade
states, respectively. The $P_{12}$ is the permutation operator for
electrons (1 $\leftrightarrow$ 2) and $P_{ij}, i,j=A,B,C$ is the
operator which interchanges the two protons $i$ and $j$. For
$S_3$-permutationally symmetric case (all protons are identical)
$\si_{\scriptstyle N}=\si_{\scriptstyle N_a}=\pm 1$. $\al_{1-6}$,
$\be_{1-2}$ and $ \gamma$ as well as $R_+ , R_-$ are variational
parameters. Their total number is eleven. It is worth emphasizing
that in the trial function (\ref{psi}) the interelectron interaction
is included explicitly in the exponential form $e^{\gamma r_{12}}$.

Calculations were performed using the minimization package MINUIT
from CERN-LIB. Multidimensional integration was carried out using a
dynamical partitioning procedure: a domain of integration was
divided into subdomains following an integrand profile and then each
subdomain was integrated separately (for details, see, e.g.,
\cite{Turbiner:2006}). Numerical integration was done with a
relative accuracy of $\sim 10^{-6} - 10^{-7}$ by use of the adaptive
D01FCF routine from NAG-LIB. A process of minimization for each
given magnetic field and for any particular state was quite
time-consuming due to a complicated profile of the total energy
surface in the parameter space but when a minimum is found it took a
few minutes to compute a variational energy.

\newpage

\section{Results}
We carry out a detailed study of $\Si, \Pi, \De$ low-lying states
with a particular emphasis of the state which has the lowest total
energy for a given magnetic field - the ground state.

\subsection{$m=0$ }
For the case $m=0$ we consider four subspaces in the Hilbert space,
$S=0$ ($m_s=0$, spin singlet states), $S=1$ (spin triplet states) at
$m_s=-1$, $\si_N=1$ (gerade states) and $\si_N=-1$ (ungerade
states).

%
%%%%%%%%%%%%%%%%%%%%%%%%%5
%
\subsubsection{$^1\Si_g$ state ($S = 0$, $\si_N =1$)}

For field-free case the system $(pppee)$ in linear configuration
(all protons are situated on a line, see, Fig.1) the state
${}^1\Si_g$ is the lowest total energy state which is characterized
by a shallow minimum (see, e.g., \cite{tennyson}). However, in spite
of developing a minimum for a linear chain the system is unstable
towards any deviation from linearity. Hence, this state is globally
unstable. It is worth noting that the true ground state does exist
and it corresponds to the equilateral triangular configuration (the
protons form equilateral triangle) with the total energy
$E_T=-2.6877\,$Ry and the side of triangle $a_{eq}=1.65\,$a.u. (see
\cite{Anderson} and also \cite{Cencek:1998})) \footnote{If the same
simple 7-parametric function (\ref{psi}) is used for the equilateral
triangular configuration it gives $E_T=-2.676\,$Ry and
$a_{eq}=1.64\,$a.u. \cite{tg} which is in quite good agreement with
the results from \cite{Anderson}}. A situation is not so different
when a magnetic field is not strong, $B \lesssim 0.2\,$a.u.: a
linear parallel configuration with the protons situated along a
magnetic line is characterized by well-pronounced minimum but a
stability towards a deviation from linearity does not occur and a
global bound state ${}^1\Si_g$ does not exist. However, with a
magnetic field growth, at $B\geq 0.2\,$a.u. the system $(pppee)$
becomes stable towards small deviations from parallel configuration
and the ${}^1\Si_g$ state exists (being an excited state, see
below).

We made a detailed study of the state ${}^1\Si_g$ of the $H_3^+$ ion
in the linear parallel configuration, with a particular emphasis of
the symmetric case $R_+ = R_- \equiv R$ (see Fig.~1), as well as
small deviations from this configuration in a wide domain of
magnetic fields $0 \le B \le 10000\,$a.u. (see Table~\ref{table1}).
Finally, for the linear parallel configuration the variational trial
function $\psi^{trial}$ (\ref{psi}) with $\si_e =1$, $\si_N =1$ and
$m=0$ was used. It depends on eleven variational parameters. A
simple, obvious generalization of (\ref{psi}) is used to study
slightly deviated configurations when stability of the linear system
was checked.

\squeezetable
\begin{table}
  \centering
\caption{
  The H$_3^+$ ion in the state  ${}^1\Si_g$ and a comparison with
  $2e$ systems H$_2$ and H$^-$:
  Total $E_T$ and binding (double-ionization) $E_I$
  energies, equilibrium distance $R_{eq}$  (in a.u.) as well as
  the total energies of final states of dissociation and ionization channels
  of H$_3^+$ are shown; all energies are in Ry. ${}^{\star}$ the
  energy and equilibrium distance of H$_3^{+}$
  for these magnetic fields is for a case when a linear
  configuration is kept externally (see text),
  {${}^{b}$}~\cite{warke}, {${}^{c}$}~our re-calculations based on the trial
  function from \cite{warke} (see text).
  Total energies for the H$_2$ molecule in ${}^1\Si_g$ state as well as H$_2^{+}$
  and H$_3^{2+}$ ions in $1\si_g$ state in a magnetic field
  taken from \cite{schmelcher}, \cite{turbiner3} and \cite{turbiner2},
  respectively. Total energies for the ground state of the H atom and for the
  H$^-$ ion in a magnetic field from \cite{turbiner} and \cite{schmelcher2},
  respectively.
  The ground state energy of H$^-$ in field-free case from \cite{pekeris}.
  }
\begin{tabular}{|c|c|c|c|c|c|c|c|c|}
\hline
     B(a.u.) & $E_T$ & $E_I$ & $R_{eq}$ &  & $E_T(H_2)$
     & $E_T(H_2^+ +H)$ & $E_T(H_3^{2+}+e)$ & $E_T(H^-)$ \\
\hline
%     &              &        &       & &         & &&\\
0$^{\star}$     &-2.5519 &    & 1.540 & & & &    -     &  \\
      &-2.55\footnote{This data can be extracted from \cite{tennyson}, p.427}
      &        &       & &         & &&\\
0.2$^{\star}$   &-2.5229 &    & 1.513 & &    &      & - &    \\
1     &-2.0692       & 4.0692 & 1.361 & & -1.7807 & -1.6122 &    -     & -0.00358\\
      &-1.7993$^{b}$ &        &       & &         &         &          &\\
      &-1.7195$^{c}$ &        & 1.402 & &         &         &          &\\
5     & 2.9597       & 7.0403 & 0.918 & &  3.6024 &         &    -     &\\
      & 3.2893$^{b}$ &        &       & &         &         &          &\\
      & 3.2888$^{c}$ &        & 0.929 & &         &         &          &\\
10    & 10.8168      & 9.1832 & 0.746 & & 11.778  & 12.1554 & 16.6084  & 15.7613\\
      & 11.154$^{b}$ &        &       & &         &         &          &\\
      & 11.153$^{c}$ &        & 0.736 & &         &         &          &\\
20    & 27.966       & 12.034 & 0.587 & &         & 30.082  &          &\\
      & 28.317$^{b}$ &        &       & &         &         &          &\\
      & 28.316$^{c}$ &        & 0.576 & &         &         &          &\\
100   & 177.59       & 22.410 & 0.336 & & 181.014 & 182.145 & 191.361  & 190.872\\
1000  & 1948.41      & 51.586 & 0.160 & &         & 1961.99  & 1979.22  & 1981.569\\
10000 & 19891.6      & 108.45 & 0.083 & &         & 19926.25 & 19954.60 &    \\
\hline
\end{tabular}
\label{table1}
\end{table}

The variational calculations demonstrate in very clear way the
existence of a minimum in the total energy surface $E_T(R_+,R_-)$
for the $(pppee)$ system for all magnetic fields ranging $B=0 -
10000$\,a.u. Minimum always corresponds to the symmetric case
$R_{eq}^+=R_{eq}^-=R_{eq}$ of the linear parallel configuration. For
$B < 0.2\,$a.u. stability is lost with respect to deviations from
linearity. This indicates a "limited" existence of the molecular ion
H$_3^+$ in the state ${}^1\Si_g$ for these magnetic fields. It
exists if in some way a linear configuration is supported
externally.

Table~\ref{table1} displays the results for the total $E_T$ and the
double ionization, $E_I=2B-E_T$, energies, as well as for the
internuclear equilibrium distance $R_{eq}$ for the state
${}^1\Si_g$. We find that with an increase of the magnetic field
strength the total energy grows more or less linearly with a
magnetic field, the system becomes more and more bound (both double
ionization and dissociation energies increase) and more compact (the
internuclear equilibrium distances $R_{eq}^{\pm}$ and a size of the
system $L_{eq}=R_{eq}^+ + R_{eq}^-$ decrease).

An important characterization of the system is given by a
description of possible dissociation and ionization channels
together with their behavior as a function of a magnetic field.
There are three dominant dissociation channels: (i) H$_3^{+} \to
\rm{H}_2 + p$, (ii) H$_3^{+} \to \rm{H}_2^+ + \rm{H}$ and (iii)
H$_3^{+} \to \rm{H}^- + p + p$ (see Table~\ref{table1}) as well as
two sub-dominant channels H$_3^{+} \to \rm{H}_2^+ + p + e$
(ionization) and H$_3^{+} \to \rm{H} + \rm{H} + p$ (dissociation).
Last two channels are characterized by higher
ionization-dissociation energies than the channel H$_3^{+} \to
\rm{H}_2^+ + $H and thus they are not considered. There are two
single-ionization processes H$_3^{+} \to \rm{H}_2^+ + p + e$ and
H$_3^{+} \to \rm{H}_3^{2+} + e$ (see Table~\ref{table1}). The second
one occurs only at $B > 10\,$a.u. where the H$_3^{2+}$ ion can
exist, it becomes a dominant single-ionization process at $B >
10000\,$a.u. where $E_T(\rm{H}_3^{2+}) < E_T(\rm{H}_2^{+})$. The
total energy of the final state compounds after dissociation for
different magnetic fields is shown in Table~\ref{table1}. It is
interesting to mention that at $B~>~100\,$a.u. the dissociation
H$_3^{+} \to \rm{H}_3^{2+} + e$ dominates over H$_3^{+} \to \rm{H}^-
+ p + p$.

A comparison of the total energy of the ground state of H$_3^+$ for
each studied magnetic field with the total energy of the products of
dissociation or ionization (see Table I) leads to a conclusion that
the total energy of the H$_3^+$ ion is always the smallest among
them. Thus, the H$_3^+$ ion in the state ${}^1\Si_g$ is stable for
all magnetic fields towards all possible dissociation or ionization
channels. A smallest dissociation energy corresponds to the channel
H$_3^{+} \to \rm{H}_2 + p$, which then is followed by H$_3^{+} \to
\rm{H}_2^+ + \rm{H}$. It is worth noting that the largest
dissociation energy corresponds to the channel H$_3^{+} \to \rm{H}^-
+ p + p$. In general, the dissociation energy (the difference
between the energies of the final and initial states) increases
monotonously with a magnetic field growth. It is quite interesting
that the difference in total energies of the final compounds of two
major dissociation channels (i) and (ii) grows extremely slow with
the magnetic field increase reaching 1.1\,Ry at $B=100\,$a.u.

A conclusion can be drawn that the H$_3^+$ molecular ion in the
state ${}^1\Si_g$ exists for $B \lesssim 0.2\,$a.u. if a linear
parallel configuration of protons is somehow supported externally,
e.g. by placing a system to an (sub)-atomic trap. However, for
larger magnetic fields it exists as an excited state which is stable
towards small deviations from linearity. It is worth noting that for
the magnetic field $B=0.2\,$a.u. the total energy well contains at
least one longitudinal vibrational state. The vibrational energy is
calculated following the same procedure which is used for H$_3^{2+}$
ion \cite{turbiner2} and it is equal to 0.035~Ry.

%\newpage

\subsubsection{${}^3\Si_u$ state ($S = 1$, $\si_N =-1$)}

%%%%%%%%%%%%%%5

In field-free case the state ${}^3\Si_u$ of the system $(pppee)$ is
(i) the only state of the H$_3^{+}$ ion in linear configuration
which is known so far and also (ii) it is the only known
spin-triplet state of H$_3^{+}$ (for a review of this state see
\cite{alijah} and references therein). For this state several
vibrational states exist. The linear symmetric configuration
$R_+=R_-$ is stable towards any small deviations, in particular,
from linearity. The state $^3\Si_u$ is stable with respect to the
decay H$_3^{+} \to \rm{H}_2^+ + \rm{H}$ (see \cite{Clementi}). Also
there is no decay channel H$_3^{+}({}^3\Si_u) \to
\rm{H}_2({}^1\Si_g) + p$.

A detailed variational study of the ${}^3\Si_u$ state of the
H$_3^{+}$ molecular ion is done for $0\,\le~B~\le~10000$\,a.u. (see
Table~\ref{Table:2}). It turns out that for all studied magnetic
fields the total energy surface displays a minimum which corresponds
to a linear parallel configuration. Furthermore, always this minimum
appears in the symmetric configuration $R_+=R_- \equiv R$. For this
particular configuration the variational trial function
$\psi^{trial}$ (\ref{psi}) with $\si_e =-1$, $\si_N =-1$ and $m=0$
is used which depends on ten variational parameters. Field-free case
is studied separately with 23-parametric trial function which is a
linear superposition of (\ref{psi}) and its three different
degenerations \footnote{Each degeneration is made in such a way that
six different $\al$'s in (\ref{psi}) are divided into three pairs
and then inside of each pair the $\al$'s are kept equal. Hence,
instead of six varying parameters $\al$'s in (\ref{psi}) we get a
degeneration where only three $\al$'s are varied.}. This
sufficiently simple function allows to reproduce three significant
digits in total energy (see Table~\ref{Table:2}). It is separately
checked that the linear parallel symmetric equilibrium configuration
is stable towards all possible small deviations.

Table~\ref{Table:2} shows the results for the total $E_T$ and the
internuclear equilibrium distance $R_{eq}$ for the ${}^3\Si_u$ state
for different magnetic fields. With an increase of the magnetic
field the total energy decreases, the system becomes more bound -
double ionization energy increases \footnote{For spin-triplet
states, $m_s=-1$ the double ionization energy is equal to
$E_I=-E_T$} and more compact (the internuclear equilibrium distance
decreases). A major emphasis of our study of the state ${}^3\Si_u$
is the domain $0.2~\lesssim B~\lesssim\,20$\,a.u. where this state
becomes the ground state of the H$_3^+$ ion in parallel
configuration and likely the global ground state of the ion.

As for the dissociation channel H$_3^{+}({}^3\Si_u) \to
\rm{H}_2^+(1\si_g) + \rm{H}(1s)$ (with electrons in spin-triplet
state) the total energy of the final state is slightly higher than
$E_T({\rm{H}_3^{+}})$ for the magnetic fields $0.2~\lesssim
B~\lesssim\,20$\,a.u.; the energy difference varies from 0.03\,Ry to
0.06\,Ry depending on a magnetic field strength, see
Table~\ref{Table:2}, remaining very small. Hence, although H$_3^{+}
({}^3\Si_u)$ is stable with respect to this dissociation channel it
turns out to be a weakly bound state. The dissociation may occur at
$B > 20\,$a.u. with photon emission at the final state. We do not
mention a dissociation channel to H$_2({}^3\Si_u) + p$ due to a
probable non-existence of the H$_2$ molecule in the domain
$0.2~\lesssim B~\lesssim\,20$\,a.u. (see e.g. \cite{schmelcher}).

\squeezetable
\begin{table}
\centering
 \caption{
 H$_3^+$ ion in the state  ${}^3\Si_u$: total energy (in Ry),
 equilibrium distance (in a.u.) and the energy of the lowest
 longitudinal vibrational state $E_0^{vib}$, rotational $E_0^{rot}$
 and bending $E_0^{bend}$. Total energy of H$_2^+(1\si_g) + \rm{H}(1s)$
 (in Ry) in ground state with spin of each electron antiparallel to
 ${\mathbf B}$ from \cite{turbiner3} and \cite{turbiner},
 respectively, shown for comparison. }
  \begin{tabular}{|c|c|c|c|c|c||c|}
\hline
B(a.u.)    & $E_T$ & $R_{eq} $ & $E_0^{vib}$ & $E_0^{rot}$ & $E_0^{bend}$
                         & $E_T(\rm{H}_2^+(1\si_g) + \rm{H}(1s))$\\
 \hline
0          & -2.2297\footnote{Our calculations (see text)} & 2.457$^{a}$ & & && -2.2052 \\
           & -2.2322\footnote{Rounded data from \cite{Clementi} and \cite{alijah}}
           & 2.454$^{b}$ & & &&         \\
0.1        & -2.3968       & 2.416       &       & &&         \\
0.2        & -2.5991       & 2.440       & 0.012 & 0.0037 & 0.014 & -2.5734 \\
0.5        & -3.0387       & 2.273       &       & &&         \\
1          & -3.6584       & 2.125       & 0.019 & 0.015  & 0.028 & -3.6122 \\
10         & -7.9064       & 1.216       & 0.048 & 0.095  & 0.17  & -7.8446 \\
20         & -10.110       & 1.00        & 0.063 & 0.16   & 0.26  & -10.082 \\
100        & -17.527       & 0.645       &       & & & -17.855 \\
1000       & -35.987       & 0.372       &       & & & -38.01  \\
10000      & -67.169       & 0.235       &       & & & -73.75  \\
\hline
  \end{tabular}
\label{Table:2}
\end{table}

In the domain $0.2~\lesssim B \lesssim~20$\,a.u. the total energy
well corresponding to the ${}^3\Si_u$ state contains at least one
longitudinal vibrational state (see Table \ref{Table:2}). Its energy
grows with a magnetic field increase. It is calculated the lowest
rotational energy as well as the lowest bending energies using the
same formulas as for H$_3^{++}$ \cite{turbiner2}. All these energies
grow with a magnetic field increase. The interesting observation is
that for each magnetic field in the domain
$1~<~B~\lesssim\,20$\,a.u. the following hierarchy of these energies
holds:
\[
   E_0^{vib} < E_0^{rot} < E_0^{bend}\ ,
\]
contrary to the hierarchy at $0.2~\lesssim B~\lesssim\,1$\,a.u.
\[
   E_0^{rot} < E_0^{vib} < E_0^{bend}\ .
\]
Hence, the bending energy is the highest to the contrary the
hierarchy at the field-free case where the longitudinal vibrational
energy is the highest (see e.g. \cite{alijah}),
\[
   E_0^{rot} < E_0^{bend} < E_0^{vib}\ .
\]

Finite-proton-mass effects might change the binding energies. So
far, it is not completely clear how such effects can be calculated
quantitatively. At present, a size of their contribution might be
estimated by values of the energies of the normal modes - the lowest
vibrational, rotational and bending energies. Their contribution to
the binding-dissociation energies grows with a magnetic field
increase (see Table II) and may reach 10-20\% for the magnetic
fields close to the Schwinger limit (for discussions and references
see \cite{Turbiner:2006}).

A comparison of the total energies of the H$_3^{+}$ ion for the
states ${}^1\Si_g$ and ${}^3\Si_u$ (see Tables~\ref{table1} and
\ref{Table:2}) shows that at $B \approx 0.2\,$a.u. the energy
crossing between these two states occurs. It implies that for linear
parallel configuration the lowest energy state for $B \lesssim
0.2$\,a.u. is the ${}^1\Si_g$ state while for $B \gtrsim 0.2$\,a.u.
the state ${}^3\Si_u$ gets the lowest total energy becoming the
ground state. Hence, one can state that in the domain $0.2~\lesssim
B \lesssim~20$\,a.u. the state ${}^3\Si_u$ is the ground state (see
below a description of $\Pi$ and $\De$ states). In this region of
the magnetic fields the linear parallel configuration for the state
${}^3\Si_u$ is stable towards small deviations. It was demonstrated
by calculating the corresponding curvatures and then the lowest
vibrational, rotational and bending energies. It is worth noting
that these energies (see Table II) turned to be small in comparison
to the $E_T=-E_I$ which implies small finite-proton-mass effects.

However, the ${}^3\Si_u$ state as a ground state is weakly bound -
energy needed for dissociation to H$_2^+(1\si_g) + \rm{H}(1s)$ with
electron spins antiparallel to ${\bf B}$ is very small. This weak
boundness can be considered as a consequence of the fact that
electrons are in the same quantum state, thus the Pauli repulsion
plays an essential role leading to a large exchange energy. It is
worth emphasizing that at $B \sim 0.2$\,a.u. the total energy of the
global ground state given by a triangular configuration \cite{tg}
coincides approximately to the total energies of the states
${}^1\Si_g$ and ${}^3\Si_u$.

\subsubsection{${}^3\Sigma_g$ state ($S = 1$, $\si_N =1 $)}

In field-free case the state ${}^3\Si_g$ of the H$_3^+$ ion in
linear configuration does not exist - the total energy surface does
not reveal a minimum or even irregularity which would correspond to
this state. However, when a magnetic field is imposed this state may
appear. It happens already at $B=1\,$a.u. where the total energy
surface $E_T(R_+,R_-)$ of this state displays a well-pronounced
minimum for linear parallel configuration. A detailed variational
study of the state ${}^3\Si_g$ of the H$_3^{+}$ molecular ion in
linear parallel configuration is done for the domain $1~\le~B~\le~
10000$\, a.u. (see Table~\ref{Table:3}). The trial function
$\psi^{trial}$ (\ref{psi}) at $\si_e =-1$, $\si_N =1$ and $m=0$ is
used, it depends on eleven variational parameters.

The calculations indicate clearly the existence of a minimum in the
total energy surface $E_T(R_+,R_-)$ of H$_3^+$ for all studied
magnetic fields $B\,=\,1 - 10000$\,a.u. The minimum always occurs
for the symmetric configuration $R_+=R_-\equiv R$. The results are
presented in Table~\ref{Table:3}. With an increase of the magnetic
field strength the total energy decreases. The system becomes more
bound: the double ionization energy $E_I$ grows [26]. Also the
system gradually becomes more compact - the internuclear equilibrium
distance gradually decreases.

\squeezetable
\begin{table}
  \centering
\caption{
  H$_3^+$ ion in the state  ${}^3\Si_g$:
  Total energy (in Ry) and equilibrium distance in (a.u.)
  (in field-free H$_3^+$ the state  ${}^3\Si_g$ does
  not exist).}
 \begin{tabular}{|c|c|c|}
\hline
B(a.u.)     &   $E_T$  & $R_{eq}$ \\
\hline
1           & -3.3256  & 5.139    \\
10          & -6.9315  & 3.063    \\
100         & -14.834  & 1.958    \\
1000        & -29.66   & 1.35     \\
10000       & -54.55   & 0.94     \\
\hline
 \end{tabular}
\label{Table:3}
\end{table}

%%%%%%%%%%%%%%%%%%%%%%%%%%%%%%%

\subsubsection{${}^1\Si_u$ state ($S = 0$, $\si_N =-1 $)}

Similar to the state ${}^3\Si_g$ in the field-free case the state
${}^1\Si_u$ of the H$_3^+$ ion in linear configuration does not
exist. However, when a magnetic field is imposed this state can
occur. Similar to the state ${}^3\Si_g$ it happens already at
$B=1\,$a.u. where the total energy surface of this state displays a
minimum. A detailed variational study of the state ${}^1\Si_u$ of
the H$_3^{+}$ molecular ion in linear parallel configuration is done
for $1\le B \le 10000$\, a.u. (see Table~\ref{Table:4}). The trial
function $\psi^{trial}$ (\ref{psi}) at $\si_e =1$, $\si_N =-1$ and
$m=0$ is used for it which depends on eleven variational parameters.

\squeezetable
\begin{table}
  \centering
\caption{
 H$_3^+$ ion in the state ${}^1\Si_u$:
      total $E_T$ and double-ionization energies $ E_I$ (in Ry), and
      equilibrium distance (in a.u.) of H$_3^+$
 (in field-free case this state does not exist). }
\begin{tabular}{|c|c|c|c|}
\hline
B(a.u.)     &  $E_T$    & $E_I$   & $R_{eq}$  \\
\hline
1           & -1.3256   & 3.3256  & 4.632   \\
10          &  13.0545  & 6.9454  & 2.563   \\
100         & 185.150   & 14.85   & 1.651   \\
1000        & 1970.36   & 29.64   & 1.494   \\
10000       & 19945.6   & 54.42   & 1.328   \\
\hline
\end{tabular}
\label{Table:4}
\end{table}

The variational calculations indicate clearly  the existence of a
minimum in the total energy surface $E_T(R_+,R_-)$ of H$_3^+$ for
magnetic fields ranging $B\,=\,1 - 10000$\,a.u. The minimum always
occurs for the symmetric configuration $R_+=R_-\equiv R$. In Table
\ref{Table:4} the results for the total $E_T$ and double ionization
energies ($E_I=2B-E_T$) as well as the internuclear equilibrium
distance $R_{eq}$ are shown. With an increase of the magnetic field
strength the total energy increases, the system becomes more bound
(double ionization energy increases) and gradually more compact (the
internuclear equilibrium distance globally decreases).

%%%%%%%%%%%%%%%%%%%%%%%%%%%%%%%%%%%%%%%%%%

\subsection{$m=-1$ }

For the case $m=-1$ four subspaces are studied: $S=0$ (spin singlet
states) and $S=1$ (spin triplet states) with $m_s=-1$, and parities
$\si_N=1$ and $\si_N=-1$, respectively. All these states do not
exist in the field-free case.

%%%%%%%%%%%%%%%%%%%%%%%%%%%%%%%%%%%%%%%%
\subsubsection{${}^3\Pi_u$ state ($S = 1$, $\si_N =1$)}

The spin-triplet state ${}^3\Pi_u$ of the H$_3^{+}$ molecular ion in
linear configuration does not exist for field-free case. However,
when a magnetic field is imposed a minimum on the total energy
surface $E_T(R_+,R_-)$ can occur. This state is studied in the
domain of magnetic fields $10^9~\le B \le 4.414 \times 10^{13}\,$G
using the variational trial function $\psi^{trial}$ (\ref{psi}) with
$\si_e=-1$, $\si_N =1$ and $m=-1$. It depends on eleven variational
parameters.

The variational calculations indicate clearly the existence of a
minimum in the total energy surface $E_T(R_+,R_-)$ of H$_3^+$ for
magnetic fields ranging $B = 1 - 4.414 \times 10^{13}\,$G. The
minimum always corresponds to a linear parallel configuration at
$R_+=R_-\equiv R$. It was investigated its stability towards small
deviation from equilibrium in linear parallel configuration. For
this state we are not aware how to check quantitatively a stability
towards deviations from linear parallel configuration. However,
physical arguments based on perturbation theory estimates indicate
to the stability. Table~\ref{Table:5} contains the results for the
total $E_T$ and the internuclear equilibrium distance $R_{eq}$. With
an increase of the magnetic field strength the total energy
decrease, the system becomes more bound (double ionization energy
increases [26]) and more compact (the internuclear equilibrium
distance decreases).

\squeezetable
\begin{table}
  \centering
\caption{
  H$_3^+$ ion for the state  $^3\Pi_u$: total energy $E_T$ (in Ry),
  equilibrium distance $R_{eq}$ (in a.u.) and the energy of the lowest
  longitudinal vibrational state $E_0^{vib}$. {$^{a}$} \cite{warke},
  {$^{b}$} Our re-calculations using the trial function from \cite{warke} (see text).
  The total energy $E_T(\rm{H}_2({}^3\Pi_u))$ is from \cite{schmelcher} for
  $B=1, 10, 100\,$a.u., while for $B=20, 1000, 10000\,$a.u. and
  $4.414 \times 10^{13}\,$G the total energy is calculated using the present
  technique (it will be described elsewhere). Data for H$_2^+(1\pi_u)$ and H$
  (1s)$ from \cite{Turbiner:2006} and \cite{Turbiner:2006He}.
  }
\begin{tabular}{|c|c|c|c|c|c|c|}
\hline B(a.u.)      &   $E_T$ & $R_{eq}$ & $E_0^{vib}$ &
                               $E_T(\rm{H}_2({}^3\Pi_u))$
            & $E_T(\rm{H}_2^+(1\pi_u) + \rm{H}(1s))$
            & $E_T(\rm{H}_2^+(1\si_g)+\rm{H}(2p_{-1}))$ \\
\hline
1           & -3.036          & 1.896    & &-2.9686 & -2.6825 & -2.8631 \\
            & -2.953  $^{a}$  &          &  &  &    &  \\
            & -2.817  $^{b}$  & 2.040    &  &  &    &  \\
5           & -5.654          & 1.163    &  &  &    &  \\
            & -5.802  $^{a}$  &          &  &  &    &  \\
            & -5.463  $^{b}$  & 1.176    &  &  &    &  \\
10          & -7.647          & 0.898    &  & -6.9325 & -6.1980 & -6.5995 \\
            & -7.803   $^{a}$ &     &  &    &     & \\
            & -7.307   $^{b}$ & 0.910    &  & &   &    \\
20          &  -9.944         & 0.706    & 0.135 & -8.934 & -8.036 & -8.582\\
            & -10.475  $^{a}$ &  & &  &          &  \\
            &  -9.752  $^{b}$ & 0.7 $^{b}$ &   &  &  & \\
100         & -18.915         &  0.395   & 0.343 & -16.473 & -14.452 & -15.547 \\
1000        & -44.538         &  0.183   & 1.105 & -35.444 & -31.353 & -33.976 \\
10000       & -95.214         &  0.093   & 3.147 & -71.39  & -62.023 & -67.356 \\
$4.414 \times 10^{13}\,$G & -115.19 & 0.078 & & -84.96 & -73.59  & -79.86 \\
\hline
\end{tabular}
  \label{Table:5}
\end{table}

The total energy of the final states for the dissociation channels
H$_3^{+} \to \rm{H}_2^+({}^1\pi_u) + \rm{H}(1s)$,  H$_3^{+} \to
\rm{H}_2^+(1\si_g)+\rm{H}(2p_{-1})$ and H$_3^{+} \to
\rm{H}_2({}^3\Pi_u) + p$ with electron spins antiparallel to the
magnetic field direction for different magnetic fields is shown in
Table~\ref{Table:5}. For all studied magnetic fields the total
energy of both dissociation channels to H$_2^+({}^1\pi_u) +
\rm{H}(1s), \rm{H}_2^+(1\si_g)+\rm{H}(2p_{-1})$ and H$_2(^3\Pi_u)$
are always higher than the total energy of the H$_3^{+}$ ion in the
${}^3\Pi_u$ state. Thus, the ion H$_3^{+}({}^3\Pi_u)$ is stable
towards these decays for all studied magnetic fields. Dominant
dissociation channel is H$_3^{+} \to \rm{H}_2({}^3\Pi_u) + p$. For
all three channels the dissociation energy grows monotonously as a
magnetic field increases. For the dominant channel H$_3^{+} \to
\rm{H}_2({}^3\Pi_u) + p$ it reaches 30.3\,Ry at the Schwinger limit
$4.414 \times 10^{13}\,$G, while for the channel H$_3^{+} \to
\rm{H}_2^+({}^1\pi_u) + \rm{H}(1s)$ for this magnetic field it is
required $\approx 35\,$Ry to dissociate. For magnetic fields $5
\times 10^{10}~\lesssim B \lesssim 4.414 \times 10^{13}\,$G there
exists at least one longitudinal vibrational state (see
Table~\ref{Table:5}).

We made an analysis of the total energies for all spin-triplet
states. One can see that there is a crossing between the ${}^3\Pi_u$
and the ${}^3\Sigma_u$ states which occurs at $B \approx 20 $\,a.u.
It shows that the ground state of H$_3^+$ for $B \gtrsim 20 $\,a.u.
is given by the ${}^3\Pi_u$ state (see below a study of $\De$ states
which are characterized by the higher total energies). While the
${}^3\Si_u$ state is the ground state for $0.2~\lesssim B \lesssim
20$\,a.u. In Figs.~2 and 3 the evolution of the total energy and the
equilibrium distance, respectively, of the ground state with the
magnetic field strength are plotted. The ground state evolves from
spin-singlet ${}^1\Si_g$ for  small magnetic fields $B\lesssim
0.2$\,a.u. (not shown in Figs.~2,3) to spin-triplet ${}^3\Si_u$ for
intermediate fields and to spin-triplet ${}^3\Pi_u$ state for $B
\gtrsim 20 $\,a.u. The total energy decreases monotonously and
smoothly as magnetic field growth. The equilibrium distance
decreases as well, but having a discontinuous behavior at $B \approx
20 $\,a.u. - in the transition from ${}^3\Si_u$ to ${}^3\Pi_u$
states. Similar behavior is displayed by $<|z_1|>$: it reduces
monotonously from $\approx 1.9\,$~a.u. at $B=0.2\,$a.u. to $\approx
0.1\,$a.u. at $B=10000\,$ a.u. with a small discontinuity at $B
\approx 20 $\,a.u. Perhaps, it is worth noting that the average
distance between two electrons $<r_{12}>$ is also reduced as a
magnetic field grows in about 20 times between 0.2 a.u. and 10000
a.u. At large magnetic fields the transverse size of the electronic
cloud coincides approximately with the Larmor radius. In Fig.~4 the
energy of the lowest longitudinal vibrational state of the ground
state for $5 \times 10^8~\lesssim B \lesssim 4.414 \times
10^{13}\,$G is presented. It grows monotonously as a magnetic field
increases suffering a discontinuity at $B \approx 20 $\,a.u. - in
the transition from ${}^3\Si_u$ to ${}^3\Pi_u$ states. In Figs.~5a,b
the valleys and the total energy behavior (profile) along the valley
for $B=100\,$a.u. for ${}^3\Pi_u$ state, respectively, are shown.
Similar behavior takes place for the valleys and the total energy
profile for ${}^3\Pi_u$ state for other magnetic fields in the
domain $5 \times 10^{10}~\lesssim B \lesssim 4.414 \times
10^{13}\,$G.

%%%%%%%%%%%%%%%%  FIGURE:2 %%%%%%%%%%%%%%%%%
%%%%%%%%%%%%%%%%%%   Total energy for all triplet states
%
%\begin{figure}[tb]
%\label{fig:energtrip}
%\begin{center}
%   \includegraphics*[width=3in,angle=-90]{energy_triplet.ps}
%    \caption{Total energies of spin-triplet states:
%       ${}^3\Pi_u$, ${}^3\De_g$, ${}^3\Si_u$,
%       ${}^3\Pi_g$, ${}^3\Si_g$ and ${}^3\De_u$.}
%\end{center}
%\end{figure}

%%%%%%%%%%%%%%%%%  Ground state energy

\begin{figure}[tb]
\begin{center}
   \includegraphics*[width=3in,angle=-90]{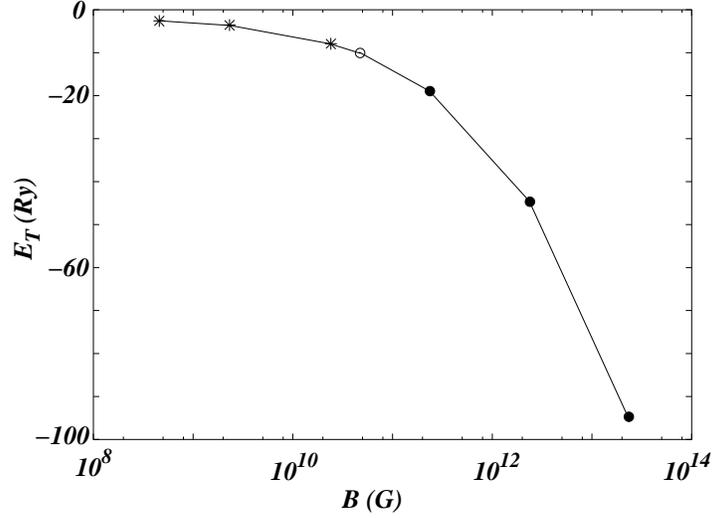}
    \caption{Ground state energy of H$_3^+$ viz magnetic field: ${}^3\Si_u$ (stars)
      and ${}^3\Pi_u$ (bullets), a point of crossing of these states is marked
      by empty circle.}
\end{center}
\label{fig:gsenergy}
\end{figure}

%%%%%%%%%%%%%%%%%  FIGURE:3 %%%%%%%%%%%%%%%%%
%%%%%%%%%%%%%%%%%  Equilibrium distance of the ground state

\begin{figure}[tb]
\begin{center}
   \includegraphics*[width=3in,angle=-90]{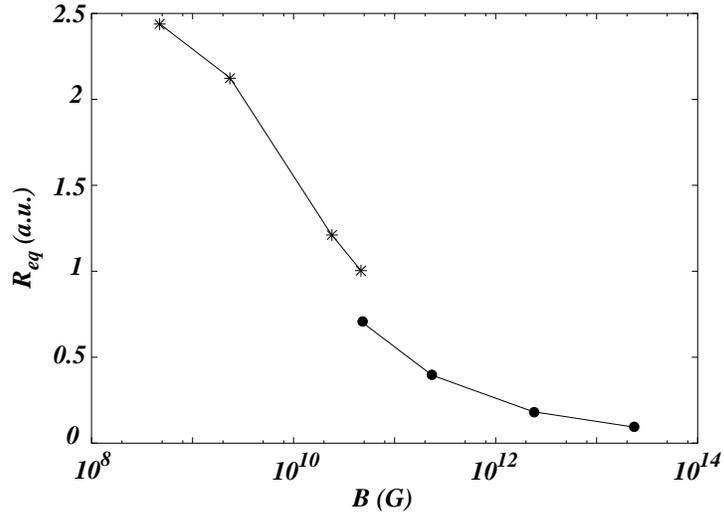}
    \caption{Equilibrium distance for the ground
      state: ${}^3\Si_u$ (stars) and ${}^3\Pi_u$ (bullets).}
\end{center}
\label{fig:distance}
\end{figure}

%%%%%%%%%%%%%%%%%  FIGURE:4 %%%%%%%%%%%%%%%%%
%%%%%%%%%%%%%%%%%  Energy  of the lowest longitudinal vibrational
%%%%%%%%%%%%%%%%%%% state Total energy for all triplet states

\begin{figure}[tb]
\begin{center}
   \includegraphics*[width=3.in,angle=-90]{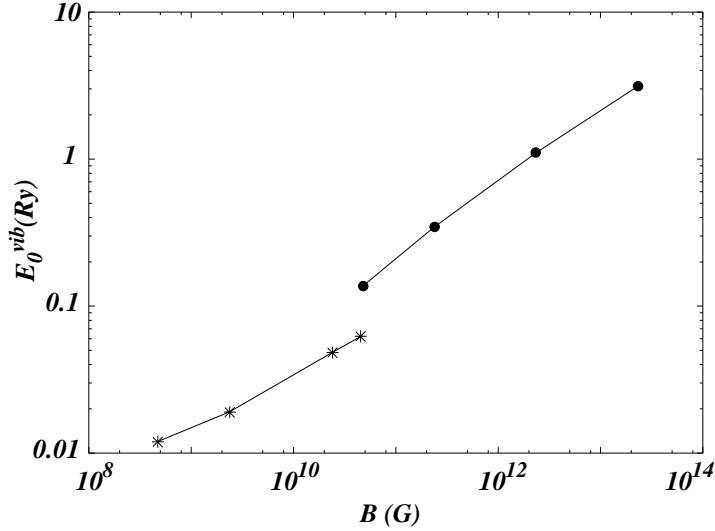}
    \caption{Energy of the lowest longitudinal vibrational state
    $E_0^{vib}$ for the ground state: ${}^3\Si_u$ (stars)
       and ${}^3\Pi_u$ (bullets). }
\end{center}
\label{fig:vibener}
\end{figure}

%%%%%%%%%%%%%%%%%  FIGURE:5 %%%%%%%%%%%%%%%%%
%%%%%%%%%%%%%%%%%  Valleys and profile %%%%%%%%%%%%%%

\begin{figure}[tb]
\begin{center}
   \includegraphics*[width=5.in,angle=0]{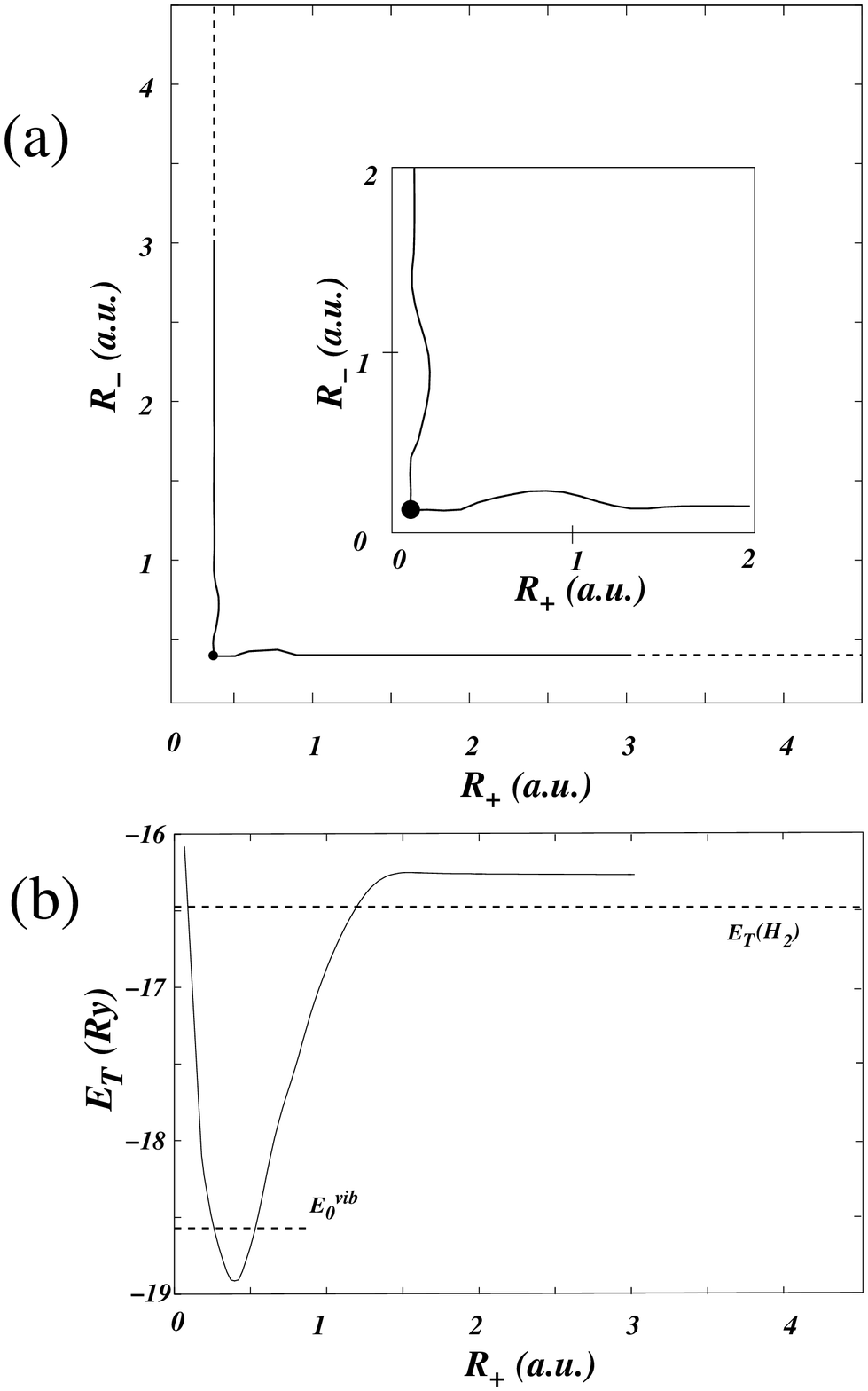}
    \caption{Valleys of the total energy (a)
       and the profile (b) for $B=100\,$a.u. of the ground state ${}^3\Pi_u$.}
\end{center}
\label{fig:valleys}
\end{figure}

\subsubsection{${}^1\Pi_u$ state ($S = 0$, $\si_N =1$)}

A detailed study of the state $^1\Pi_u$ of the H$_3^{+}$ molecular
ion in symmetric configuration $R_+=R_-\equiv R$ is carried out in
the domain of magnetic fields $1~\le~B~\le~10000\,$a.u. (see
Table~\ref{Table:6}). The variational trial function $\psi^{trial}$
with $\si_e =1$, $\si_N =1$ and $m=-1$ is used for this state, it
depends on ten variational parameters.

The obtained results indicate clearly the existence of a minimum in
the total energy $E_T(R)$ of H$_3^+$ for all magnetic fields ranging
$B=1 - 10000$\,a.u. Table \ref{Table:6} shows the total $E_T$ and
double ionization energies ($E_I=2B-E_T$), as well as the
internuclear equilibrium distance $R_{eq}$ for the ${}^1\Pi_u$
state. It is found that with an increase of the magnetic field
strength the total energy increases, the system becomes more bound
(double ionization energies increase) and more compact (the
internuclear equilibrium distance decreases).

\begin{table}
  \centering
\caption{
 H$_3^+$ ion in the state  ${}^1\Pi_u$: total $E_T$ and double-ionization
 $E_I$ energies (in Ry) and equilibrium distance $R_{eq}$ (in a.u.).
 ${}^{a}$ \cite{warke}, ${}^{\rm b}$~our
 re-calculations using the trial function from \cite{warke} (see text). }
\begin{tabular}{|c|c|c|c|}
\hline
B(a.u.)     &  $E_T$          &  $ E_I$   &  $R_{eq}$ \\
\hline
1           & -0.809         &  2.809    &   1.995   \\
            & -0.561  $^{a}$ &           &           \\
            & -0.511  $^{b}$ &           &   2.233   \\
5           &  4.747         &  5.253    &   1.232   \\
            &  5.025  $^{a}$ &           &           \\
            &  5.051  $^{b}$ &           &   1.262   \\
10          & 13.028         &  6.972    &   0.967   \\
            & 13.346  $^{a}$ &           &           \\
            & 13.346  $^{b}$ &           &   0.963   \\
20          & 30.708         &  9.292    &   0.750   \\
            & 31.078  $^{a}$ &           &           \\
            & 31.081  $^{b}$ &           &   0.738   \\
100         & 182.23         &  17.77    &   0.419   \\
1000        & 1957.77        &  42.23    &   0.191   \\
10000       & 19909.0        &  91.0     &   0.098   \\
\hline
\end{tabular}
\label{Table:6}
\end{table}

%%%%%%%%%%%%%%%%%%%%%%%%%%%%%%%%%%%%%%%%%%%5
\subsubsection{${}^1\Pi_g$ state ($S = 0$, $\si_N =-1$)}

It is carried out a detailed study for the state  ${}^1\Pi_g$ of the
H$_3^{+}$ molecular ion in symmetric configuration $R_+=R_-\equiv R$
in the domain of magnetic fields
$1~\mbox{a.u.}~\le~B~\le~10000\,$a.u. (see Table~\ref{Table:7}). For
this state our variational trial function $\psi^{trial}$ with
$\sigma_e =1$, $\sigma_N =-1$ and $m=-1$ depends on ten variational
parameters. The total $E_T$ and double ionization $E_I=2B-E_T$
energies increase while the internuclear equilibrium distance
$R_{eq}$ decreases as a magnetic field grows; the system becomes
more bound (double ionization energies increase) and more compact
(the internuclear equilibrium distance decreases).

\begin{table}
  \centering
\caption{
  H$_3^+$ ion in the state  ${}^1\Pi_g$: total energy $E_T$ in Ry
  and equilibrium distance $R_{eq}$ in a.u.}
 \begin{tabular}{|c|c|c|c|}
\hline
B(a.u.)     &  $E_T$     & $E_I$    & $R_{eq}$ \\
\hline
1           &    -0.701  &  2.701  &  3.176   \\
10          &    13.669  &  6.331  &  1.441   \\
100         &   185.413  & 14.587  &  0.741   \\
1000        &  1969.42   & 30.58   &  0.421   \\
10000       & 19942.1    & 57.9    &  0.273   \\
\hline
 \end{tabular}
  \label{Table:7}
\end{table}

%%%%%%%%%%%%%%%%%%%%%%%%%%%%%%%%%%%%%%5

\subsubsection{${}^3\Pi_g$ state ($S = 1$, $\si_N =-1$)}

A detailed study is carried out for the state  ${}^3\Pi_g$ of the
H$_3^{+}$ molecular ion in symmetric configuration $R_+=R_-\equiv R$
in the domain of magnetic fields
$1~\mbox{a.u.}~\le~B~\le~10000\,$a.u. (see Table~\ref{Table:8}). For
this state our variational trial function $\psi^{trial}$ (\ref{psi})
with $\si_e =-1$, $\si_N=-1$ and $m=-1$ depends on ten variational
parameters. The total $E_T$ energy decreases and the double
ionization $E_I$ energy [26] increases while the internuclear
equilibrium distance $R_{eq}$ decreases as a magnetic field grows;
the system becomes more bound (double ionization energy increases)
and more compact (the internuclear equilibrium distance decreases).

\begin{table}
  \centering
\caption{ H$_3^+$ ion in the state ${}^3\Pi_g$:
  total energy $E_T$ in Ry and equilibrium distance $R_{eq}$ in a.u. }
\begin{tabular}{|c|c|c|}
\hline
B(a.u.)     &  $E_T$    &  $R_{eq}$ \\
\hline
1           &  -2.6095  &  2.700    \\
10          &  -6.276   &  1.487    \\
100         & -14.429   &  0.838    \\
1000        & -30.44    &  0.447    \\
10000       & -57.8     &  0.27    \\
\hline
\end{tabular}
\label{Table:8}
\end{table}

\subsection{$m=-2$ }

In the $m=-2$ subspace we study four subspaces: $S=0$ (spin singlet
states), $S=1$ (spin triplet states with $m_s=-1$),  $\si_N=1$  and
$\si_N=-1$ and the lowest energy state in each of them. All these
states do not exist in the field-free case.

It is carried out a detailed study for the states of the symmetric
configuration $R_+=R_-\equiv~R$ in the domain of magnetic fields
$1~\le~B~\le~10000\,$a.u. For each of these four states ${}^1\De_g,
{}^3\De_g, {}^1\De_u, {}^3\De_u$ the trial function $\psi^{trial}$
(\ref{psi}) at $m=-2$ depends on ten variational parameters. All
four states indicate clearly the existence of a minimum in the total
energy $E_T(R)$ of H$_3^+$ for magnetic fields ranging $B=1 -
10000$\,a.u. Tables~\ref{Table:9} - \ref{Table:12} show the results.
For these states with an increase of the magnetic field strength the
total energy increases for the spin-singlet states and decreases for
spin-triplet states, the system becomes more bound (double
ionization energy increases) and more compact (the internuclear
equilibrium distance decreases).
\begin{table}[h]
  \centering
\caption{ The H$_3^+$ ion for the state  $^1\Delta_g$:
  total $E_T$ and double-ionization $E_I$ energies in Ry and
  equilibrium distance $R_{eq}$ in a.u. }
\begin{tabular}{|c|c|c|c|}
\hline
$B$\, (a.u.)   & $E_T$   &  $ E_I$   &   $R_{eq}$ \\
\hline
1           &   -0.6136 &  2.6136  &  2.206   \\
10          &   13.499  &  6.501   &  1.027   \\
100         &  183.325  & 16.675   &  0.433   \\
1000        & 1960.19   & 39.81    &  0.191   \\
10000       & 19913.6   & 86.4     &  0.10  \\
\hline
\end{tabular}
\label{Table:9}
\end{table}
%%%%%%%%%%%%%%%%%%%%%%%%%%%%%%%%%%%%%%%%%%%%%%%%%%%%%%%%%%5
\begin{table}[h]
  \centering
\caption{The H$_3^+$ ion in the state  $^3\De_g$:
   total energy $E_T$ in Ry and equilibrium distance
   $R_{eq}$ in a.u. }
\begin{tabular}{|c|c|c|}
\hline
B(a.u.)     &   $E_T$    & $R_{eq}$ \\
\hline
1           &   -2.633   &  2.179  \\
10          &   -6.624   &  1.013  \\
100         &  -16.92    &  0.432  \\
1000        &  -40.38    &  0.197  \\
10000       &  -87.49    &  0.099  \\
$4.414 \times 10^{13}\,$G & -106.02 & 0.09 \\
\hline
\end{tabular}
\label{Table:10}
\end{table}
%%%%%%%%%%%%%%%%%%%%%%%%%%%%%%%%%%%%%%%
\begin{table}[h]
  \centering
\caption{The H$_3^+$ ion in the state  ${}^1\De_u$:
    total energy $E_T$ in Ry and equilibrium distance
    $R_{eq}$ in a.u.}
\begin{tabular}{|c|c|c|c|}
\hline
B(a.u.)     & $E_T$     & $ E_I$   &  $R_{eq}$ \\
\hline
1           &   -0.4107 &   2.4107 &  3.316    \\
10          &   14.281  &   5.719  &  1.514    \\
100         &  186.602  &  13.398  &  0.775    \\
1000        &  1972.08  &  27.92   &  0.401    \\
10000       & 19945.7   &  54.3    &  0.273    \\
\hline
\end{tabular}
\label{Table:11}
\end{table}
%%%%%%%%%%%%%%%%%%%%%%%%%%
\begin{table}[h]
  \centering
\caption{The H$_3^+$ ion in the state  ${}^3\De_u$:
   total energy $E_T$ in Ry and equilibrium distance
   $R_{eq}$ in a.u. }
\begin{tabular}{|c|c|c|}
\hline
B(a.u.)     & $E_T$   & $R_{eq}$ \\
\hline
1           & -2.443  & 4.494   \\
10          & -5.722  & 1.600   \\
100         & -13.39  & 0.804   \\
1000        & -28.41  & 0.449   \\
10000       & -54.4   & 0.28    \\
\hline
\end{tabular}
\label{Table:12}
\end{table}

\section{Conclusion}

We study the low-lying energy states of H$_3^+$ molecular ion in
linear configuration parallel to a magnetic field from 0  up to
$4.414 \times 10^{13}\,$G using the variational method in the
Born-Oppenheimer approximation. The total energy curves display a
well pronounced minimum at finite internuclear distances at
$R_+=R_-$ for the lowest states with magnetic quantum numbers
$m=0,-1,-2$, total spins $S=0,1(m_s=-1)$ and parity $p=\pm 1$. A
level distribution for several magnetic field strengths is shown on
Fig.~6. If in field-free case there exist only two eigenstates in a
linear configuration, but many more states in linear parallel
configuration can appear when a magnetic field is imposed.

%%%%%%%%%%%%%%%%%  FIGURE:6 %%%%%%%%%%%%%%%%%
%%%%%%%%%%%%%%%%%  Energy of the lowest excited states

\begin{figure}[htb]
\begin{center}
   \includegraphics*[height=9.5in,width=7.in,angle=0]{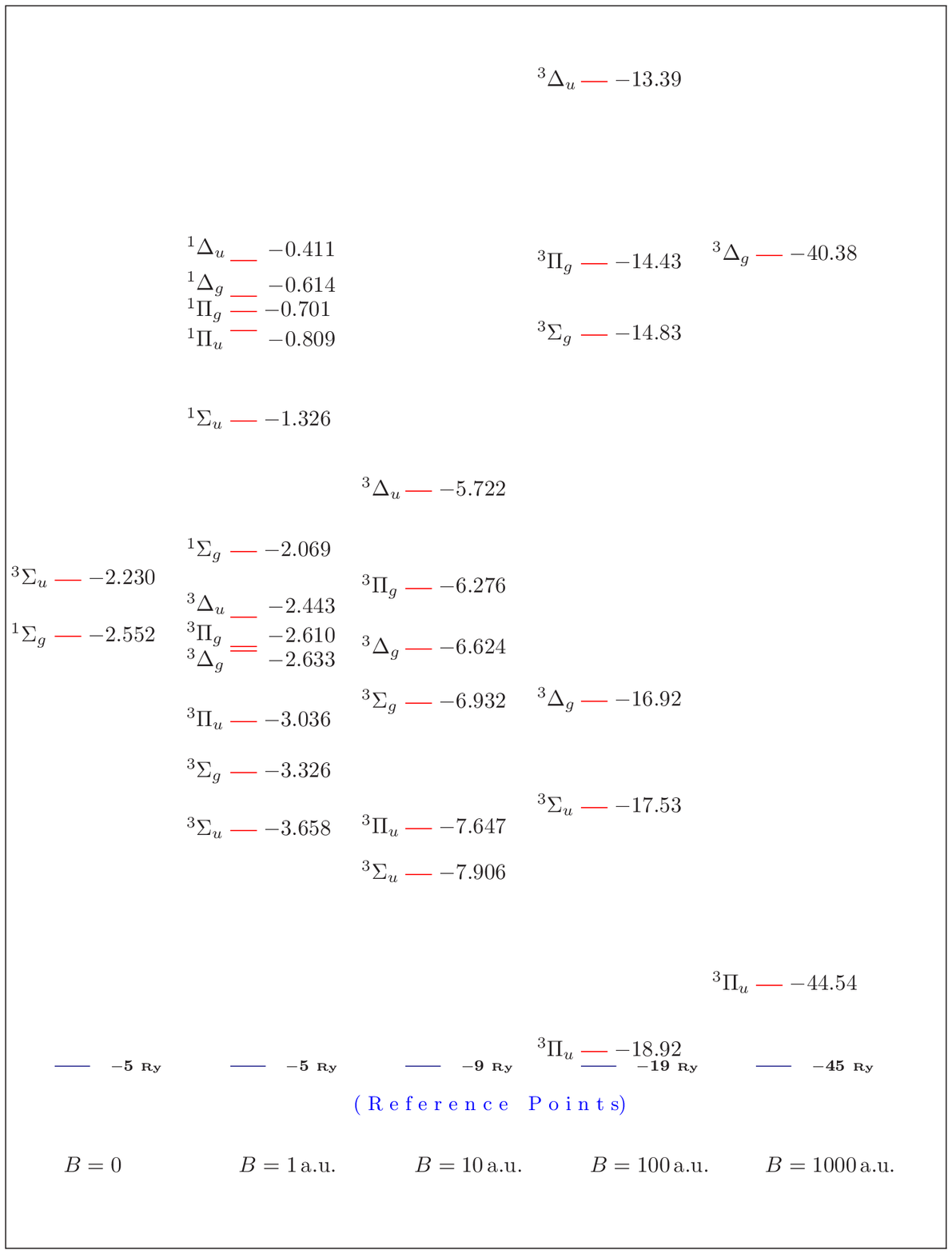}
   \vskip -70pt
    \caption{Total energy of  the low-lying levels for
    $B=0, 1, 10, 100\ \mbox{and}\ 1000\,$ a.u.
    (energy scale is kept the same for all presented magnetic fields but the
     reference points depend on them)}
\end{center}
\label{fig:levels}
\end{figure}

In general, for all studied states, as the magnetic field increases
the equilibrium internuclear distances $R_{eq}$ decreases and the
system becomes more compact, while the total energies of
spin-singlet states increase whereas that of spin-triplet states
decrease.

The state of the lowest total energy in linear parallel
configuration depends on the magnetic field strength. It evolves
from spin-singlet (unstable towards a deviation from linearity)
$^1\Si_g$ for weak magnetic fields $B\lesssim 0.2\,$a.u. to
spin-triplet (stable towards a deviation from linearity) ${}^3\Si_u$
for intermediate fields and eventually to spin-triplet ${}^3\Pi_u$
state for $B \gtrsim 20 $\,a.u. which remains the ground state until
the Schwinger limit $B = 4.414 \times 10^{13}\,$G.  It is worth
emphasizing that for weak magnetic fields, $B\lesssim 0.2\,$a.u.,
the global ground state is given by a triangular configuration
\cite{tg} and then, for larger magnetic fields, the global stable
ground state corresponds to a linear parallel configuration
\footnote{In order to make such a claim that the state of the lowest
energy corresponds by a linear parallel configuration we make a very
natural physically assumption that there are no any other spacial
configuration which may provide a lower total energy. But in order
to be rigorous we must investigate the total energy surface for all
possible spacial configurations.}. The H$_3^+$ ion in the
${}^3\Si_u$ state is weakly bound. For all studied magnetic fields
the total energy surface well corresponding to the ground state
contains at least one longitudinal vibrational state.

It is interesting to compare the evolution of the ground state for
H$_3^+$ with magnetic field change with that of other two-electron
systems (see \cite{Turbiner:2006London} and references therein). For
atomic type H$^-$ and He systems there is no domain of magnetic
field where the spin-triplet, $m=0$ state is the ground state: for
weak fields the ground state is the spin-singlet, $m=0$ state and
then it becomes the spin-triplet, $m=-1$ state for large fields. For
the hydrogen molecule the ${}^3\Si_u$ state is unbound for all
magnetic fields unlike the case of H$_3^+$. It implies that the
H$_2$ molecule does not exist as a bound system for $0.18 \lesssim B
\lesssim 15.6$\,a.u., where the unbound state ${}^3\Si_u$ has the
lowest total energy at infinitely-large distance between protons. A
similar situation occurs for the He$_2^{2+}$-ion: it does not exist
as a bound system for $0.85 \lesssim B \lesssim 1100$\,a.u.
\cite{Turbiner:2006He2}.

What is the lowest-lying excited state for weak magnetic fields
$B\lesssim 0.2\,$a.u. is not clear yet. This question, and also the
whole domain $B\lesssim 0.2\,$a.u., will be studied elsewhere. In
the domain of magnetic fields $0.2~\le~B~\le~5\,$a.u. the
lowest-lying excited state is ${}^3\Si_g$, then for $B \gtrsim
5\,$a.u. the lowest-lying excited state is ${}^3\Pi_u$. For $B
\gtrsim 20 \,$a.u., where the ${}^3\Pi_u$ state becomes the ground
state, the lowest-lying excited state is ${}^3\Si_u$. However, at $B
\gtrsim 1000 \,$a.u. until the Schwinger limit the lowest-lying
excited state is ${}^3\De_g$.

It is interesting to note that at $B = 1000\,$a.u. the H$_3^+$ ion
exists with ${}^3\Pi_u$ as the ground state ($E_T=-44.54\,$a.u.)
with two possible excited states: ${}^3\De_g$ ($E_T=-40.38\,$a.u.)
and ${}^3\Si_u$ ($E_T=-35.99\,$a.u.) with energies below the
threshold of dissociation to H$_2({}^3\Pi_u) + p$
($E_T=-35.44\,$a.u.). For larger magnetic fields the situation
becomes different. For instance, at $B = 10000 \,$a.u. for the
H$_3^+$ ion ($E_T=-95.21\,$a.u.) only one excited state, ${}^3\De_g$
($E_T=-87.45\,$a.u.), exists with energy below the dissociation
threshold to H$_2({}^3\Pi_u) + p$ ($E_T=-71.39\,$a.u.). Similar
situation holds up to the Schwinger limit $B=4.414 \times
10^{13}\,$G: a single excited state ${}^3\De_g$ lies below the
dissociation threshold.

It is found that many states in linear configuration which do not
exist for $B=0\,$ begin to be bound at relatively small magnetic
field $B \approx 0.2\,$a.u. A study of the existence of the bound
states which might appear in a spacial configuration is our goal for
a future study. Another goal is related to a study of transition
amplitudes for different electronic states.

Present consideration is based on the use of a simple variational
trial function (\ref{psi}). This function can be easily generalized
and extended in the same way as was done in a variational study of
various one-electron systems in a strong magnetic field (see
\cite{Turbiner:2006}). This will allow to improve the present
results and might be done in future. However, we are not sure that
such a study is crucially important. It is related to a fact that
typical accuracies in astronomical observations of neutron star
radiation would not be higher $10^{-3} - 10^{-4}$ unlike to
spectroscopical accuracies in laboratory where they can be by
several orders of magnitude higher.

\begin{acknowledgments}
 The first author is grateful to A.~Alijah, T.~Oka and J.~Tennyson for
 introduction to the subject of the H$_3^+$ ion, helpful discussions
 and interest to the present work. We thank D. Page for careful
 reading of the manuscript.
 Computations were performed on a dual DELL PC with two Xeon processors
 of 2.8\,GHz each (ICN), 54-node FENOMEC and 32-node TOCHTLI clusters (UNAM)
 and a dual DELL PC with two Xeon processors of 3.06\,GHz each (CINVESTAV).
 This work was supported in part by CONACyT grant
 {\bf 47899-E} and PAPIIT grant {\bf IN121106} (Mexico).
\end{acknowledgments}

%%%%%%%%%%%%%%%%%%%%%%%%%%%%%%%%%%%%%%%%%%%%%%%%%%%%
%%%%%%%%%%%%%Bibliography%%%%%%%%%%%%%%%%%%%%%%%%%%%
%%%%%%%%%%%%%%%%%%%%%%%%%%%%%%%%%%%%%%%%%%%%%%%%%%%%

\end{document}